\documentclass{aastex}
\usepackage{emulateapj5}

\def\ts     {\thinspace}
\def\kms    {\ifmmode{{\rm \ts km\ts s}^{-1}}\else{\ts km\ts s$^{-1}$}\fi}
\def\msol     {\ifmmode{{\rm M}_{\odot}}\else{M$_{\odot}$}\fi} 
\def\etal   {{\rm et\ts al.}}
\def\aco  {\ifmmode{^{12}{\rm CO}(J\!=\!1\! \to \!0)}\else{$^{12}{\rm CO}(J\!=\!1\! \to \!0)$}\fi}
\def\bco  {\ifmmode{^{12}{\rm CO}(J\!=\!2\! \to \!1)}\else{$^{12}{\rm CO}(J\!=\!2\! \to \!1)$}\fi}
\def\m    {\ifmmode{\mu {\rm m}}\else{$\mu$m}\fi}
\def\cco  {\ifmmode{^{13}{\rm CO}(J\!=\!1\! \to \!0)}\else{$^{13}{\rm CO}(J\!=\!1\! \to \!0)$}\fi}
\def\dco  {\ifmmode{^{13}{\rm CO}(J\!=\!2\! \to \!1)}\else{$^{13}{\rm CO}(J\!=\!2\! \to \!1)$}\fi}
\def\eco  {\ifmmode{{\rm C}^{18}{\rm O}(J\!=\!1\! \to \!0)}\else{${\rm C}^{18}{\rm O}(J\!=\!1\! \to \!0)$}\fi}
\def\nh   {\ifmmode{N(\hi)}\else{$N$(\hi)}\fi}
\def\hun    {\ifmmode{I_{100}}\else{$I_{100}$}\fi}
\def\sex    {\ifmmode{I_{60}}\else{$I_{60}$}\fi}
\def\hh     {\ifmmode{{\rm H}_2}\else{H$_2$}\fi}
\def\nhh     {\ifmmode{N({\rm H}_2)}\else{$N$(H$_2$)}\fi}
\def\zwco   {\ifmmode{^{12}{\rm CO}}\else{$^{12}{\rm CO}$}\fi}
\def\nzwco   {\ifmmode{N(^{12}{\rm CO})}\else{$N(^{12}{\rm CO})$}\fi}
\def\wzwco   {\ifmmode{W(^{12}{\rm CO})}\else{$W(^{12}{\rm CO})$}\fi}
\def\drco   {\ifmmode{^{13}{\rm CO}}\else{$^{13}{\rm CO}$}\fi}
\def\ndrco   {\ifmmode{N(^{13}{\rm CO})}\else{$N(^{13}{\rm CO})$}\fi}
\def\wdrco   {\ifmmode{W(^{13}{\rm CO})}\else{$W(^{13}{\rm CO})$}\fi}
\def\tex    {\ifmmode{T_{ex}({\rm CO})}\else{$T_{ex}({\rm CO})$}\fi}
\def\ha     {\ifmmode{{\rm H}\alpha}\else{${\rm H}\alpha$}\fi}
\def\amm     {\ifmmode{{\rm NH}_{3}}\else{${\rm NH}_{3}$}\fi}
\def\xco     {\ifmmode{X_{\rm CO}}\else{$X_{\rm CO}$}\fi}

\newcommand{\hi}{H\,{\small{\sc I}}}
\newcommand{\ci}{C\,{\small{\sc I}}}
 
\slugcomment{}

\shorttitle{First Detection of Ammonia in M\,82}
\shortauthors{Wei\ss\ et al.}

\begin{document}

\title{First Detection of Ammonia in M\,82}

\author{A. Wei\ss\ }
\affil{Radioastronomisches Institut der Universit\"at Bonn, 
Auf dem H\"ugel 71, 53121 Bonn, Germany}
\email{aweiss@astro.uni-bonn.de}

\author{N. Neininger}
\affil{Radioastronomisches Institut der Universit\"at Bonn, 
Auf dem H\"ugel 71, 53121 Bonn, Germany}

\author{C. Henkel}
\affil{ Max--Planck--Institut f\"ur Radioastronomie, 
Auf dem H\"ugel 69, 53121 Bonn, Germany}
\author{J. Stutzki}
\affil{I. Physikalisches Institut der Universit\"at zu K\"oln, Z\"ulpicher 
Stra{\ss}e 77, 50937 K\"oln, Germany}

\and

\author{U. Klein}
\affil{Radioastronomisches Institut der Universit\"at Bonn, 
Auf dem H\"ugel 71, 53121 Bonn, Germany}

\begin{abstract}
We report the detection of the (J,K) = (1,1), (2,2), and (3,3)  
inversion lines of ammonia (\amm) towards the south--western molecular  
lobe in M\,82. The relative intensities of the ammonia lines are  
characterized by a rotational temperature of $T_{rot} = 29\pm5$\,K which  
implies an average kinetic temperature of $T_{kin}\approx 60\,$K.  
A Gaussian decomposition of the observed spectra indicates increasing 
kinetic temperatures towards the nucleus of M\,82, consistent with recent 
findings based on CO observations.  The observations imply a very low  
\amm\ abundance relative to \hh, $X(\amm) \approx 5\times 10^{-10}$. 
We present evidence for a decreasing \amm\ abundance towards the central 
active regions in M\,82 and interpret this abundance gradient in terms
of photodissociation of \amm\ in PDRs. The low temperature derived 
here from \amm\ also explains the apparent underabundance of complex
molecules like CH$_3$OH and HNCO, which has previously been reported.
\end{abstract}


\section{Introduction}
The physical properties of molecular clouds are important parameters 
which are believed to influence largely the rate at which stars can form 
by gravitational collapse of cloud fragments. The nearby starburst  
galaxy M\,82 is known to host a large amount of molecular gas  
($M(\hh)\approx 1\times10^8\msol$, e.g. Wild \etal\ \cite{wild}) and 
is currently 
undergoing a phase of intense star 
formation. Therefore it has been a prime target to determine the physical  
conditions of the molecular phase of its ISM in order  
to increase our understanding of the relevant physical processes  
connected with violent star formation activity.\\ 
Most of these studies use CO emission lines to derive the physical   
properties of the molecular gas (e.g. Harris \etal\ \cite{harris};   
Wild \etal\ \cite{wild}, and more recently Mao \etal\ \cite{mao}; 
Petitpas \& Wilson \cite{petitpas}; Wei\ss\ \etal\ \cite{weiss}). Yet the  
interpretation of CO emission lines requires the application of radiative  
transfer models which use simplifying assumptions about the structure and 
kinematics of molecular clouds. Therefore it is desirable to observe other  
molecules which directly trace the physical conditions and thus allow a 
test of conclusions gained from CO observations. Ammonia (\amm) 
is such a molecule: the relative population of its meta--stable inversion 
lines is sensitive to the kinetic gas temperature (e.g. Walmsley \&  
Ungerechts \cite{walmsley}). Therefore observations of ammonia inversion  
lines can be used as a molecular cloud thermometer. Here we report the  
first detection of ammonia towards the most prominent molecular 
source in M\,82, the south--western molecular lobe. We compare our 
interpretation of the ammonia lines to conclusions deduced from CO 
observations.

\section{Observations}
The (J,K) = (1,1), (2,2), (3,3) and (4,4) 
inversion lines of ammonia were observed towards the south--western 
(SW) molecular lobe in M\,82 
($\alpha_{2000}=09^h55^m49.06^s$, $\delta_{2000} =69^\circ40'41''.2$) using  
the Effelsberg 100\,m telescope of the MPIfR equipped with a dual channel 
K--band HEMT receiver. The system temperature was 180--200\,K on a 
main beam brightness temperature ($T_{\rm mb}$) scale. The 
observations were carried out in May 2000 and Feb. 2001. The beam 
size of the 100-m telescope at 23\,GHz is $\approx 40''$. The 
data were recorded using an autocorrelator with $8\times512$ channels 
and a bandwidth of 40\,MHz for each backend, leading to a channel 
spacing of 1\,\kms.  The eight backends  
were configured such that all four ammonia lines were observed  
simultaneously. Each pair of backends centered on a given ammonia line 
was used to sample both linear polarizations independently. To avoid a  
contamination of the data by low--level baseline instabilities we shifted 
the center velocity of the backends (May 2000) and permuted the backends 
(Feb. 2001) every 90 minutes.\\
The measurements were carried out in dual--beam switching mode, 
with a switching frequency of 1Hz and a beam throw of $2'$ in azimuth.  
Flux calibration was obtained by observing both the continuum and 
the ammonia lines of W3(OH) before each observing run (for fluxes see 
Mauersberger, Wilson, \& Henkel \cite{mauer88} and Ott \etal\ \cite{ott}). 
Pointing was checked every 1.5 hours on the nearby continuum source 
0836+71 and was found to be stable to within $5-10''$.
A linear baseline was removed from each spectrum and intensities were 
converted to a $T_{\rm mb}$ scale. The summed spectra were smoothed to a 
velocity resolution of 16\,\kms. We estimate the flux calibration of the 
final reduced spectra to be accurate within $\pm20$\% (10\% error of the 
flux calibration and 10\% uncertainty due to low--level baseline  
instabilities). 

\section{Results \label{results}}
\begin{figure}[h] 
\hspace*{0cm} 
\resizebox{8.5cm}{!}{\includegraphics{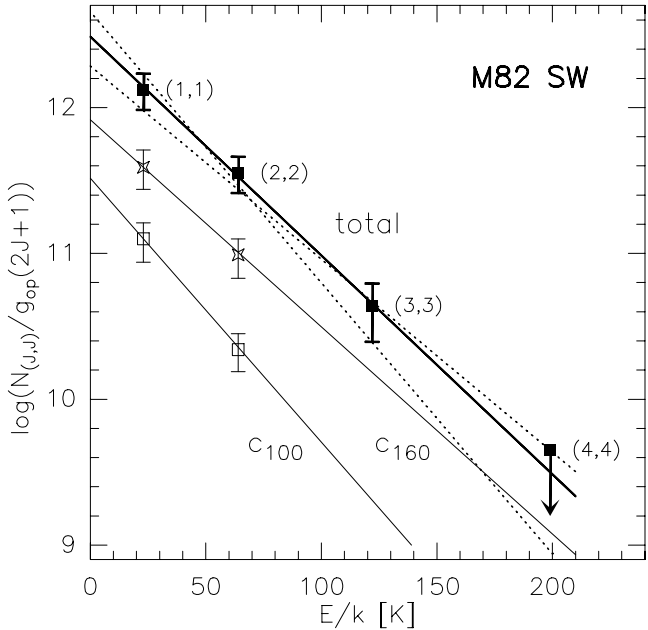}} 
\caption{Rotation diagram of meta--stable ammonia transitions towards the 
SW molecular lobe in M\,82. The filled squares show the  
normalized column densities determined from the integrated line intensities.  
The thick solid line (denoted by 'total') corresponds to a linear fit to the  
(J,K) = (1,1) and (2,2) lines for these values. The dotted lines  
correspond to the linear fits with the lowest and highest slope, which are  
still consistent with the data including the upper limit for the  
(J,K) = (4,4) line. The open squares 
and the stars show the normalized column densities determined from a Gaussian 
decomposition of the (J,K) = (1,1) and (2,2) spectra into the velocity  
components C$_{100}$ and C$_{160}$. Note that 
for display purpose the values for C$_{160}$ and C$_{100}$ have been 
shifted by --0.2 and --0.8 on the y--axis scale, respectively. The thin
solid lines correspond to linear fits to these data points.} 
\label{nh3-trot} 
\end{figure}

The observed spectra of the ammonia lines are shown in Fig.\,\ref{nh3-lines}. 
The (J,K)\,=\,(1,1)--(3,3) lines are detected with a S/N ratio better than 4.  
The (4,4) line is not detected. The (1,1) and (2,2) lines peak at
$v_{lsr}=100\,\kms$ (C$_{100}$) and show a weaker component at 
$v_{lsr}=160\,\kms$ (C$_{160}$). Both velocity components are 
also detected in low-- and mid--J CO emission lines (see e.g. Harris \etal\ 
\cite{harris}; Wild \etal\ \cite{wild}; Mao \etal\ \cite{mao};  
Petitpas \& Wilson \cite{petitpas}). At $40''$ resolution the intensity 
of the low--J CO transitions from C$_{160}$ is stronger than that from 
C$_{100}$. To emphasize this point we show the line profile of
 the \aco\ transition at $40''$ resolution in the top panel of 
Fig.\,\ref{nh3-lines}. \\ 
Beam averaged column densities for individual inversion states were  
calculated using 
\begin{equation} 
\nonumber N(J,K) = \frac{7.77\cdot10^{13}}{\nu}  
\frac{J(J+1)}{K^2}\int T_{mb} dv 
\end{equation} 
(e.g. Henkel \etal\ \cite{henkel2000}). The column density $N$, the frequency  
$\nu$ and the integrated line intensity are in units of cm$^{-2}$, GHz and  
K\,\kms, respectively. Line parameters and column densities are summarized  
in Tab.\,\ref{line-para}. Note that this approximation  
assumes optically thin emission and the contribution of the 2.7\,K  
background to be  negligible (${\rm T}_{ex} \gg 2.7\,{\rm K}$). Following 
the analysis described by Henkel \etal\ (\cite{henkel2000}) the  
rotation temperature ($T_{rot}$) between levels J and J' can be determined 
from the slope of a linear fit in the rotation diagram (normalized column  
density vs. energy above the ground state expressed in E/k) by 
\begin{equation} 
 T_{rot}  = \frac{-{\rm log}({\rm e})}{a} = \frac{-0.434}{a}  
\end{equation} 
where $a$ is the slope of the linear fit.\\ 
The rotation diagram for the observed ammonia lines is shown in  
Fig.\,\ref{nh3-trot}. The rotation temperature between the (1,1) and (2,2)  
inversion levels of para--ammonia is $T_{rot} = 29\,{\rm K}$ (thick  
solid line). The uncertainty, derived from the extrema of the slope  
including the (4,4) level as an upper limit, is $\pm5\,{\rm K}$ 
(dotted lines). The (J,K)\,=\,(3,3) line is not included in the fit because  
it belongs to ortho--ammonia, thus to a different ammonia species.  
Nevertheless, it nicely fits to the rotation temperature derived  
from the two lowest inversion levels of para--ammonia.\\ 
In addition to the analysis of the integrated line intensities we have 
decomposed the (J,K)\,=\,(1,1) and (2,2) spectra into two Gaussian 
components with fixed center velocities of $v_{lsr}=100\,\kms$ (C$_{100}$)  
and 160\,\kms (C$_{160}$). The Gaussian decomposition is shown 
together with the observed line profile in the two top panels  
of Fig.\,\ref{nh3-lines}. A separate analysis of the rotation  
temperatures between the (1,1) and (2,2) inversion levels for both  
components yields $T_{rot,100}=24_{-6}^{+12},{\rm K}$ and  
$T_{rot,160}=31_{-10}^{+22}\,{\rm K}$ for C$_{100}$ and C$_{160}$,  
respectively. For the errors we have assumed 30\% uncertainty from the  
Gaussian decomposition. Note that the observed  
upper limit for the (J,K)\,=\,(4,4) line intensity is still consistent with 
the shallow slope derived for the emission arising from the 160\,\kms\  
component (C$_{160}$).     
  
\section{Discussion}

\subsection{The \amm\ emitting volume}
\begin{figure}[h] 
\hspace*{0cm} 
\resizebox{8.5cm}{!}{\includegraphics{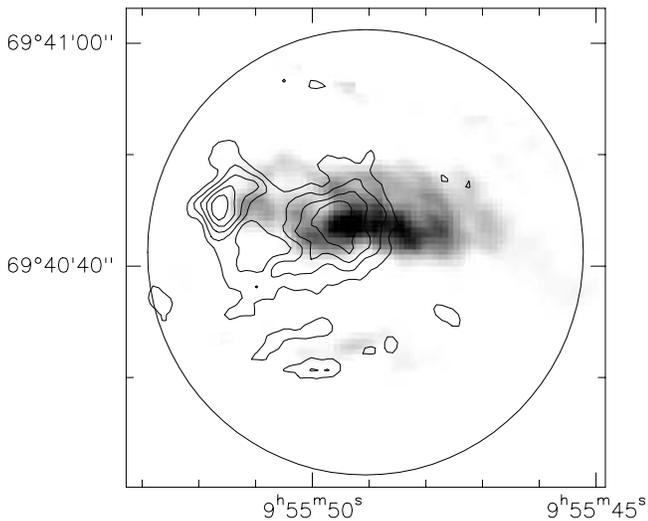}} 
\caption{High--spatial resolution \bco\ data integrated over 
$v_{lsr}=130-190\,\kms$ (${\rm C}_{160}$, contours) juxtaposed on \bco\  
data integrated over $v_{lsr}=80-120\,\kms$ (${\rm C}_{100}$, greyscale).  
The $40''$ beam size of the Effelsberg telescope is indicated by the circle.} 
\label{co-highres} 
\end{figure}

High--spatial resolution CO observations show that the emission at  
$v_{lsr}\approx 100\,\kms$ is mainly associated with the SW  
molecular lobe. The emission at $v_{lsr}\approx 160\,\kms$ arises     
from regions closer to the nucleus , i.e. from the central 
molecular peak and the inner CO outflow region. Thus C$_{160}$ covers 
the regions of the western mid infrared peak close to the central 
molecular peak (Telesco \& Gezari \cite{telesco92}), and the 
region where an expanding molecular superbubble  was identified 
(Wei\ss\ \etal\ \cite{weiss99}). Therefore  C$_{160}$  
represents a more active region than the outer molecular lobe itself.  
An overlay of the two regions at high spatial resolution ($\approx 2''$)  
as observed in the \bco\ transition (Wei\ss\ \etal\ \cite{weiss}) 
is shown in  Fig.\,\ref{co-highres}. 

\subsection{Comparison with temperature determinations from 
other line observations \label{comp}}

Radiative transfer calculations (e.g. Walmsley \& Ungerechts \cite{walmsley}) 
of \amm\ show that rotation temperatures determined from meta--stable levels 
only reflect the kinetic gas temperature for low ($T<15\,$K)  
temperatures. For larger temperatures the rotation temperature largely  
underestimates the kinetic gas temperature due to depopulation  
mechanisms. Correcting our results for these effects, our mean  
rotation temperature of 29\,K corresponds to a mean kinetic  
temperature of $T_{kin}\approx60\,$K.  
Using the Gaussian decomposition we find $T_{kin}\approx45\,$K for the  
temperature in the SW molecular lobe (C$_{100}$), and  
$T_{kin}\approx80\,$K for the regions closer to the nucleus (C$_{160}$).\\ 
These values are in good agreement with recent high--resolution  
kinetic temperature estimates using radiative transfer calculations based  
on CO observations by Wei\ss\ \etal\ (\cite{weiss}). They  
derive $T_{kin}\approx55\,$K for the SW molecular lobe, and  
an average of $T_{kin}\approx110\,$K over the region which corresponds to  
C$_{160}$ in our study. The simultaneous observations of two [\ci] 
fine structure lines towards the SW lobe allowed 
Stutzki \etal\ (\cite{stutzki1997}) to derive a lower limit for the 
kinetic temperature of the [\ci] emitting gas in C$_{100}$ of 
50\,K which is consistent with the temperature derived above 
from the \amm\ lines. Similar values ($T_{kin}=50\,$K) have been 
found by Seaquist \& Frayer (\cite{seaquist}) using HCO${^+}$ and HCN 
emission lines.\\ 
Note, however, that the uncertainty of the kinetic 
temperature estimates from the Gaussian decomposition of the \amm\ spectra  
is quite large. Nevertheless, the results are in line with the general  
picture that the kinetic temperature rises from the outer parts of the  
SW molecular lobe towards the active regions closer to the 
nucleus of M\,82.\\ 
In comparison with rotation temperatures determined from \amm\ in IC\,342  
($T_{rot} \approx  50\,{\rm K}$; Martin \& Ho \cite{martin}) and Maffei~2  
($T_{rot} \approx  85\,{\rm K}$; Henkel \etal\ \cite{henkel2000}) the values 
determined here for M\,82 are very low. Thus the temperature of the dense 
molecular gas in M\,82 is obviously much lower than in other 
galaxies with large nuclear concentrations of molecular gas. 
This explains the apparent 
underabundance of molecules like SiO, CH$_3$OH, HNCO, and CH$_3$CN in M\,82 
which only form in a dense and {\it warmer} environment (see e.g.  
Mauersberger \& Henkel \cite{mauer93}).

\subsection{\amm\ abundance}
\begin{figure}[ht] 
\hspace*{0cm} 
\resizebox{8.3cm}{!}{\includegraphics{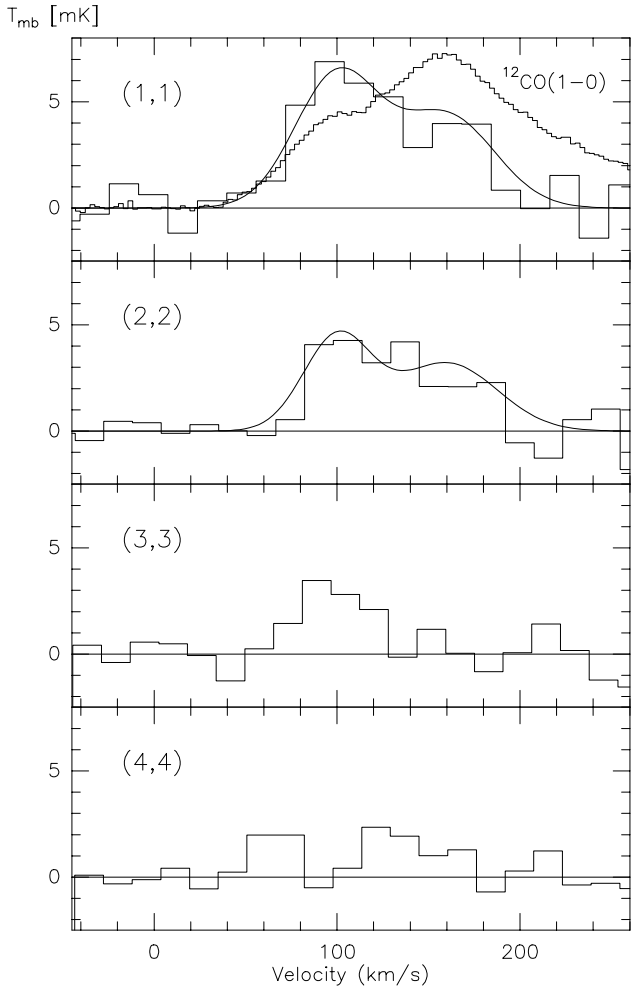}} 
\caption{\amm\ spectra towards the SW part of M\,82  
($\alpha_{2000}=09^h55^m49.06^s$, $\delta_{2000} =69^\circ40'41''.2$). 
The velocity resolution is 16\,\kms\ for each spectrum. The high--resolution 
spectrum in the top panel shows the \aco\  
emission line at the same position and  spatial resolution ($40''$) scaled 
by 1/370. For the fits to the (1,1) and (2,2) lines see Sect.\,3.} 
\label{nh3-lines} 
\end{figure}

\begin{figure}[ht] 
\hspace*{0cm} 
\resizebox{8.5cm}{!}{\includegraphics{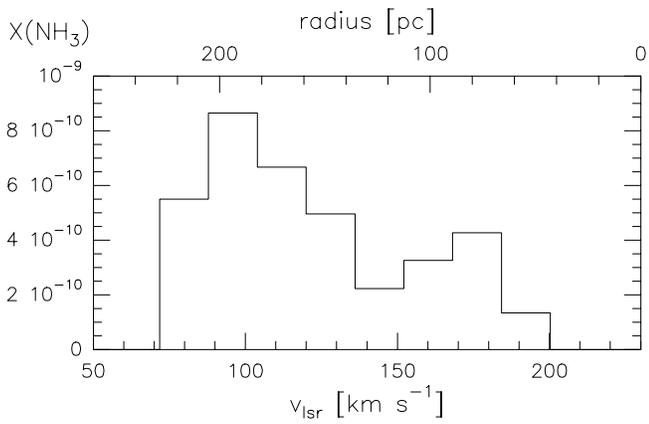}} 
\caption{\amm\ abundance relative to \hh\ per velocity interval. The upper 
axis denotes the distance of the emission region from the nucleus of M\,82 
assuming pure solid body rotation and a velocity gradient of  
0.7\,\kms\,pc$^{-1}$.} 
\label{nh3-abun} 
\end{figure}

Using eq. A15 of Ungerechts, Walmsley, \& Winnewisser (\cite{ung1986}) we  
have estimated a total beam--averaged \amm\ column density of 
$N(\amm) \approx 1\times10^{13}\,{\rm cm}^{-2}$. The beam--averaged 
\hh\ column density was estimated to be $\nhh 
\approx 1.9\times10^{22}\,{\rm cm}^{-2}$ using the CO spectrum shown 
in Fig.\,\ref{nh3-lines} and a conversion factor of $\xco =
5\times10^{19}\,{\rm cm^{-2} \,(K \kms)^{-1}}$, which corresponds 
to the average value of the conversion factors derived by Wei\ss\ \etal\  
(\cite{weiss}) in the region covered by 100--180 \kms\ emission. 
This yields a relative abundance of ammonia of  
$X(\amm) \approx 5\times 10^{-10}$, which is an extremely low value. 
In nearby dark clouds the fractional \amm\ abundance 
is of order $X(\amm) \approx 10^{-7}$ (e.g. Benson \& Myers \cite{benson}) 
and in hot cores \amm\ is even more abundant ($X(\amm) \approx 10^{-5...-6}$; 
e.g. Mauersberger, Henkel \& Wilson \cite{mauer87}).
In a recent work on Maffei~2 Henkel \etal\ (\cite{henkel2000}) found 
$X(\amm) \approx 10^{-8}$.\\ 
A closer look at the (J,K)\,=\,(1,1) and the \aco\ spectra 
displayed in Fig.\,\ref{nh3-lines} (top) reveals another interesting aspect 
regarding the spatial variation of the \amm\ abundance: 
the line temperature ratio $T(\amm(1,1))/T(\aco)$ decreases with  
increasing velocity for $v_{lsr} > 100 \kms$. The solid--body rotation 
in the inner part of M\,82 allows one to associate velocities lower than 
the systemic velocity of $v_{lsr} = 230 \kms$ with a specific distance  
from the nucleus (e.g. Neininger \etal\ \cite{nico}). 
Therefore, $T(\amm(1,1))$ per velocity interval traces the ammonia  
column density in a region much smaller than the spatial resolution,
while $T(\aco)$ traces the \hh\ column density. Note that conversion 
from I(CO) to \nhh\ changes across the major axis of M\,82 and therefore 
for each velocity interval (Wei\ss\ \etal\ \cite{weiss}). 
Fig.\,\ref{nh3-abun} shows a histogram with $X(\amm)$ versus radial 
velocity and galactocentric radius, accounting for variations of \xco\ as 
determined by Wei\ss\ \etal\ (\cite{weiss}). The \amm\ abundance is found 
to decrease towards the center of M\,82. We believe that this 
finding reflects a real change of the ammonia abundance, because a 
constant \amm\ abundance would imply that $\xco$ changes by a factor of more
than 40 on a linear scale of 300\,pc which is not consistent with radiative 
transfer models. Due to its low energy threshold 
for photodissociation ($\approx\,4.1$\,eV, Suto \& 
Lee \cite{suto}), \amm\ should be destroyed rapidly in PDRs (G\"usten \& 
Fiebig \cite{g+f}). This process should be even more efficient 
when the bulk of the gas is distributed in a diffuse phase with low \hh\ 
column densities as it seems to be true for the central regions in M\,82 
(Mao \etal\ \cite{mao}, Wei\ss\ \etal\ \cite{weiss}). In such an environment
shielding against the strong UV radiation is ineffective which leads to 
low \amm\ abundances towards the central star forming regions.
Note, that \amm\ abundances are not always small in a warm environment. 
Efficient release of \amm\ into the gas phase by evaporation of dust grain 
mantles is known to occur in galactic `hot cores'. This process, however, 
becomes efficient at slightly higher temperatures than those obtained by us 
for NH$_3$ in Sect.\,\ref{comp}.\\
We therefore interpret the abundance gradient as a result of ammonia 
being almost completely dissociated in the harsh environment close 
to M\,82's nucleus, whereas it can still partially survive, with low 
abundance, in the cooler, denser, and thus better shielded, clouds of the 
SW lobe.

\acknowledgments{This research project was supported by the 
Deutsche Forschungsgemeinschaft grant SFB~494.}

\newpage

\clearpage

\begin{table*}[t] 
\caption{Parameters of the ammonia lines towards the SW molecular 
lobe of M\,82. Column 2 to 6 correspond to the line parameters
of the entire line profiles derived from a visual inspection. Columns 7 and 8
give the column densities for the (1,1) and (2,2) transitions derived
from a gaussian decomposition of the spectra.}
\begin{tabular}{c l l l l l l l} 
\hline 
\hline 
Transition & $T_{mb}^{peak}$ & $\int T_{mb} dv$ & $V^{peak}_{LSR}$ &  
$\Delta V_{1/2}$ & $N(J,K)$& $N(J,K)_{100}$& $N(J,K)_{160}$ \\ 
(J,K) & (mK) & (K\,\kms) & (\kms) & (\kms) & ($10^{12}$ cm$^{-2}$)
& ($10^{12}$ cm$^{-2}$)& ($10^{12}$ cm$^{-2}$)\\  
\hline 
(1,1) & $6.9\pm0.8$ & $0.61\pm0.16$ & $97\pm16$  & $94\pm11$  & $4.0\pm1.1$ 
&$2.2\pm0.8$&$1.8\pm0.6$\\ 
(2,2) & $4.3\pm0.9$ & $0.36\pm0.11$ & $105\pm16$ & $80\pm10$  & $1.8\pm0.5$ 
&$0.9\pm0.3$&$1.0\pm0.3$\\ 
(3,3) & $3.5\pm0.7$ & $0.14\pm0.06$ & $95\pm10$  & $47\pm20$  & $0.6\pm0.26$
&--&--\\ 
(4,4) & $<2.1\,(3\sigma)$ & $<0.1$ &--&--&$<0.4$&--&--\\ 
\hline 
\end{tabular} 
\label{line-para} 
\end{table*}

\end{document}